# Absence of ferroelectricity in BiMnO$_3$ ceramics

V. Goian,[1] S. Kamba,[1,*] M. Savinov,[1] D. Nuzhnyy,[1] F. Borodavka,[1] P.Vaněk,[1] A. A. Belik[2]

[1]*Institute of Physics, Academy of Sciences of the Czech Republic, Na Slovance 2, 18221 Prague 8, Czech Republic*
[2]*International Center for Materials Nanoarchitectonics (WPI-MANA), National Institute for Materials Science (NIMS), 1-1 Namiki, Tsukuba, Ibaraki 305-0044, Japan*

We performed factor-group analysis of all phonons in possible monoclinic *C2/c* and *C2* structures of BiMnO$_3$ and compared it with our experimental infrared and Raman spectra. We conclude that the crystal structure is centrosymmetric *C2/c* in the whole investigated temperature range from 10 to 550 K, therefore BiMnO$_3$ cannot be ferroelectric. We revealed a dielectric relaxation in THz spectra above the structural phase transition taking place at T$_{C1}$=475 K giving evidence in strong lattice anharmonicity and a large dynamical disorder of Bi cations above T$_{C1}$. Step-like dielectric anomaly observed at T$_{C1}$ in THz permittivity reminds antiferroelectric phase transition. Nevertheless, the low-temperature dielectric studies did not reveal any antiferroelectric or ferroelectric hysteresis loop. Our experimental results support theoretical paper of P. Baettig et al. (J. Am. Chem. Soc. **129**, 9854 (2007)) claiming that BiMnO$_3$ is not multiferroic, but only antipolar ferromagnet.

## I. Introduction

BiMnO$_3$ was first time described already in 1960's,[1] but it is intensively studied mainly in the last decade after theoretical prediction of Hill (Spaldin) and Rabe[2] that this system could be ferroelectric ferromagnet, because some of the zone center phonons should be unstable according to their calculations. Most of multiferroics are ferroelectric antiferromagnets and have a small magnetoelectric coupling. Simultaneous coexistence of ferromagnetic and ferroelectric order can enhance the magnetoelectric coupling and therefore the BiMnO$_3$ becomes subject of intensive studies (see review of Belik[3]).

There is no doubt in literature about the ferromagnetic order in BiMnO$_3$, it was confirmed by many researches. Curie temperature was determined between 99 and 102 K[4,5,6,7] and the saturated magnetization reaches 3.9 μ$_B$ at 5 K, which is close to the expected value of 4.0 μ$_B$. Magnetization is oriented along the monoclinic *b*-axis.[8] Ferroelectricity in BiMnO$_3$ is still subject of many controversial research studies. In thin films, some papers reported about large room-temperature (RT) remnant polarizations ($P_r$=9-16 µC/cm$^2$)[9], other authors found only $P_r$=0.004 - 0.03 µC/cm$^2$ at low temperatures around 120 K,[10,11] some researches did not confirm any ferroelectricity.[3] The same controversial results were obtained in BiMnO$_3$ ceramics. Two papers[10,12] found ferroelectric hysteresis loops with the $P_r$ values of 0.043 µC/cm$^2$ at 200 K[10] and 0.06 µC/cm$^2$ at RT[12]. However, the hysteresis loops were not saturated, which can indicate the loops come from dielectric losses. There is also no general agreement about the RT crystal structure of BiMnO$_3$. Some reports claim non-centrosymmetric *C2* structure,[13,14] other newer papers propose centrosymmetric *C2/c* structure,[15,16,17] which rules out ferroelectricity in this system. Nevertheless, it seems that the centrosymmetric *C2/c* structure can transform to non-centrosymmetric *C2* structure after irradiation of the bulk samples or after induction of some defects (e.g. oxygen vacancies) in BiMnO$_3$ lattice.[3,12] Some thin films of BiMnO$_3$ are ferroelectric and their structure is non-centrosymmetric probably due to the non-stoichimetry or large applied electric field. Newest reports show that the RT *C2/c* structure of bulk BiMnO$_3$ is stable from liquid He temperatures up to $T_{C1}$ = 474 K, above which the structure changes to another *C2/c* phase.[1,15,18,4] Only at $T_{C2}$ = 760-770 K the structure of BiMnO$_3$ changes to a GdFeO$_3$ type phase with *Pnma* space group,[18,4] but no details are known about this structure, because the BiMnO$_3$ system decomposes above 770 K.[4]

The nature of the phase transition at $T_{C1}$ is mysterious. Long time was believed that $T_{C1}$ is the ferroelectric phase transition temperature from *C2/c* to *C2* space group. Large dielectric anomaly was observed at $T_{C1}$[19], but this anomaly was not confirmed in other papers. The anomaly probably comes from Maxwell-Wagner polarization, which arises from electric conductivity of the ceramics and it is known that the conductivity of BiMnO$_3$ changes exactly at $T_{C1}$.[18] Although the recent investigations claim that the symmetry does not change at $T_{C1}$, pronounced anomalies were observed in X-ray diffraction, lattice constants etc.[18,20]

The question arises how will be demonstrated the structural phase transition near $T_{C1}$ in infrared (IR) and Raman spectra. If the phase transition would be proper and displacive ferroelectric, a ferroelectric soft mode should be observed. If the phase transition is improper ferroelectric, some phonons should disappear in high-temperature phase due to change of selection rules. In the case of order-disorder ferroelectric phase transition, some dielectric relaxation in paraelectric phase and peak in permittivity should be observed at $T_{C1}$. Simultaneously some polar phonons should disappear above $T_{C1}$ due to transition to centrosymmetric paraelectric phase. If the symmetry does not change at $T_{C1}$, the selection rules for phonon activities do not change, i.e. no phonons should disappear above $T_{C1}$ and only some small shifts of phonon frequencies are expected.

We will show in this paper that the phonons observed in IR and Raman spectra of BiMnO$_3$ can be explained within centrosymmetric monoclinic *C2/c* structure. Absence of ferroelectric or antiferroelectric hysteresis loop supports theoretically predicted[25] antipolar order in BiMnO$_3$.

**II. Experimental**

The BiMnO$_3$ ceramics were prepared using high-purity Bi$_2$O$_3$ and Mn$_2$O$_3$ powder under 6 GPa in a belt-type high-pressure apparatus at 1383 K for 60-70 min as described elsewhere.[5,15]

Low-frequency (1 Hz – 1 MHz) dielectric measurements were performed between 10 and 300 K using NOVOCONTROL Alpha-A High Performance Frequency Analyzer. The ferroelectric hysteresis loops were measured at frequencies of 1-50 Hz and temperatures between 10 and 150 K, at higher temperatures the ceramics were too conducting.

For the terahertz (THz) time-domain transmission experiments we used a Ti:sapphire femtosecond laser oscillator. Linearly polarized THz probing pulses were generated by an optical rectification in [110] ZnTe crystal plate and detected using the electro-optic sampling with a 1 mm thick [110] ZnTe crystal. The complex dielectric spectra were taken in the range 5-50 cm$^{-1}$ (150 GHz – 1.5 THz) at temperatures from 10 to 550 K.

Near-normal IR reflectivity spectra were obtained using a Fourier transform IR spectrometer Bruker IFS 113. The IR measurements were performed up to 550 K using a commercial high-temperature sample cell SPECAC P/N 5850. The same cell was used also for high-temperature THz experiment in nitrogen atmosphere. In both THz and IR measurements, Optistat CF cryostat (Oxford Instruments) with polyethylene (IR) and Mylar (THz) windows were used for measurements between 10 and 300 K. Frequency range of the low-temperature IR measurements was limited by transparency of polyethylene windows (up to 650 cm$^{-1}$), the measurements above room temperature were performed up to 3000 cm$^{-1}$, which allowed us to determine not only phonon parameters but also $\varepsilon_\infty$ i.e. the sum of electronic contributions to permittivity.

Near-normal IR reflectivity can be expressed as

$$R(\omega) = \left| \frac{\sqrt{\varepsilon^*(\omega)} - 1}{\sqrt{\varepsilon^*(\omega)} + 1} \right|^2 \qquad (1)$$

where complex dielectric function $\varepsilon^*(\omega)$ can be expressed using the sum of damped quasiharmonic oscillators and one Debye relaxation

$$\varepsilon^*(\omega) = \varepsilon_\infty + \sum_{j=1}^{n} \frac{\Delta\varepsilon_j \omega_{TOj}^2}{\omega_{TOj}^2 - \omega^2 + i\omega\gamma_j} + \frac{\Delta\varepsilon_R \omega_R}{\omega_R + i\omega} \qquad (2)$$

where $\omega_{TOj}$, $\Delta\varepsilon_j$ and $\gamma_j$ are the eigenfrequency, dielectric strength and damping of the $j^{th}$ polar phonon, respectively. $\Delta\varepsilon_R$ and $\omega_R$ are dielectric strength and relaxation frequency of the Debye relaxation, respectively. The static permittivity $\varepsilon(0)$ is given by the sum of all contributions

$$\varepsilon(0) = \sum_{j=1}^{n} \Delta\varepsilon_j + \Delta\varepsilon_R + \varepsilon_\infty \qquad (3)$$

Eqs. (1) – (2) were used for simultaneous fits of IR reflectivity and THz permittivity.

For high-temperature Raman studies, a Renishaw RM 1000 Micro-Raman spectrometer equipped with a CCD detector and a Linkam THMS 600 temperature cell was used.

Thin ceramic $BiMnO_3$ plates (m = 84.5 mg) placed in aluminum pan were measured in a differential scanning calorimeter Perkin Elmer Pyris Diamond DSC using control and evaluation software Pyris 4.02. Sample was heated and cooled three times in the temperature range from 300 to 553 K by a rate of 10 K/min. Nitrogen was used as a purging gas.

**III. Results and discussions**

Broad-band dielectric permittivity as well as results of hysteresis loops measurements are shown in Fig. 1. The low-temperature permittivity has intrinsic value around 30 because it corresponds, within accuracy of our measurements, to the sum of phonon and electronic dielectric strengths (see Eq. (3) and Fig. 1(a)). The conductivity of ceramic grains rapidly increases from ~$10^{-8}$ to ~$10^{-5}$ S/cm on heating from 20 to 300 K, while the grain boundaries have conductivity ~$10^{-13}$ S/cm at 20 K which rapidly increases on heating above 100 K and reaches values ~ $10^{-5}$ S/cm near room temperature (not shown). From that reason tan$\delta$ increases above 1 and permittivity reaches extrinsic giant values > $10^3$ at temperatures above 100 K. This is well known Maxwell-Wagner polarization mechanism giving rise to giant extrinsic permittivities at high temperatures.[21] Vibration of ferroelectric domain walls usually contributes to permittivity, but it was not observed in our case. We found no significant dielectric dispersion between 1 Hz and 335 GHz below 100 K (see Fig. 1(a)).

We tried to measure as well the ferroelectric hysteresis loops (see Figure 1b), but without success. Above 100 K an ellipsoid typical for lossy dielectrics was obtained. At lower temperatures only paraelectric behavior is seen. No signature of ferroelectricity was discovered, our $BiMnO_3$ ceramics behaves like typical paraelectric. We note that ferroelectric hysteresis loops in bulk $BiMnO_3$ ceramics were reported in Refs. [10,12], but the observed polarizations were below 0.1 μC/cm$^2$ and the loops were never saturated. These could be dielectric lossy loops, whose ellipsoids can become two artificial sharp heads due to the triangular-shaped increase and decrease of the applied voltage.[22] Sinusoidal change of voltage was used in our case.

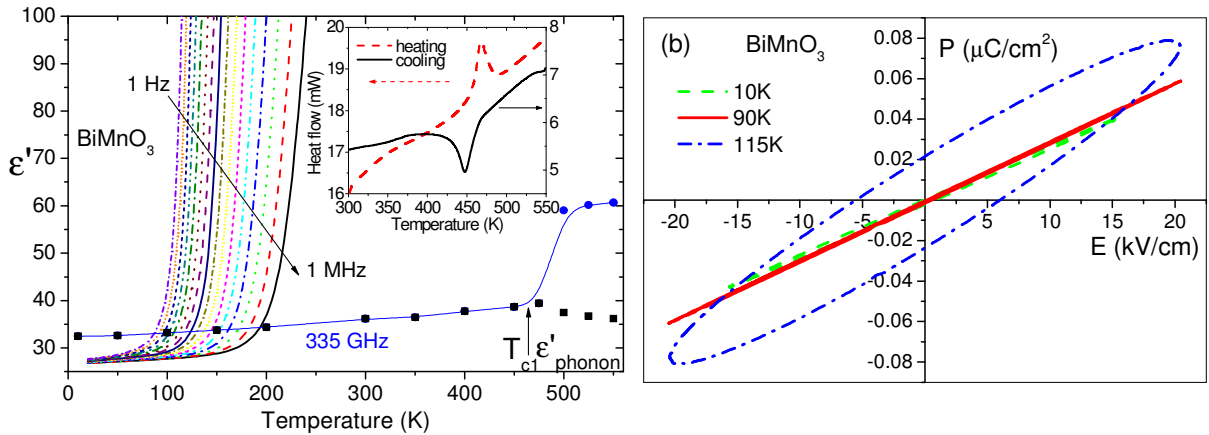

FIG. 1. (Color online) (a) Temperature dependence of permittivity at various frequencies. (b) Polarization versus electric field plotted at various temperatures show paraelectric behavior below 90 K and lossy loop at 115 K. In inset of Figure 1(a) the heat flow behavior measured on heating and cooling with a rate of 10 K/min is shown. The full black squares in (a) figure show sum of phonon contributions to permittivity, while the blue line and solid points (ε' measured at 335 GHz) take into account also THz dielectric relaxation seen only above $T_{C1}$.

We tried to prove whether the previously reported structural phase transition takes place at $T_{C1}$ in our $BiMnO_3$ ceramics. Calorimetry measurements presented in inset of Figure 1a clearly reveal a heat anomaly near 470 K (taken on heating). On cooling the anomaly is observed 30 K lower, which gives evidence for first-order character of the phase transition. The observed change of enthalpy was 1.6 J/g at $T_{C1}$. We did not heat up the ceramics above

550 K, because we were afraid of the chemical decomposition, which occurs near $T_{C2}$=760 K.[4]

Infrared reflectivity spectra plotted at selected temperatures up to 550 K are shown in Fig. 2(a). On heating one can see gradual decrease of reflection band intensities due to increase of phonon damping. Thanks to this fact, some phonons become overlapped and therefore they disappear from the spectra at high temperatures. Two lowest-frequency phonons exhibit small softening towards $T_{C1}$ (Fig. 3), leading to a small maximum of phonon permittivity at $T_{C1}$ (see black solid squares in Fig. 1a). Nevertheless, the most remarkable change undergo low-frequency THz spectra showing abrupt increase of both $\varepsilon'$ and $\varepsilon''$ above $T_{C1}$ (see blue solid circles in Figs. 2(b) and 2(c)). This can be fitted only by dielectric relaxation (Eq. (1)), which expresses a dynamical disorder of some atoms above $T_{C1}$. The relaxation frequency slows-down from 4 cm$^{-1}$ (i.e. 120 GHz) at 550 K to 2.8 cm$^{-1}$ (84 GHz) at 500 K and finally it disappears below the structural phase transition at $T_{C1}$=475 K. The observed slowing down of dielectric relaxation gives evidence in order-disorder mechanism of the structural phase transition at $T_{C1}$, but the relaxation frequency in THz range is rather unusual. Most of materials with order-disorder mechanism of the ferroelectric phase transition exhibit relaxation frequency in microwave or radio-frequency region. On the other hand, dielectric relaxation only in THz region is rather rare. It was observed for example above antiferroelectric phase transition in Nd substituted $BiFeO_3$ [23] as well as in antiferroelectric $PbZrO_3$.[24] Our discovered step down of $\varepsilon'(T)$ seen on cooling at $T_{C1}$ (see Fig. 1a) reminds an antiferroelectric phase transition. We did not find any antiferroelectric hysteresis loop in Fig. 1b, but it can be a consequence of relatively low applied electric field 20 kV/cm (the sample becomes leaky at higher voltage). Possible antipolar order in $BiMnO_3$ was theoretically proposed by Baetig et al.,[25] who used LDA+U method of density functional theory. They found that the $Bi^{3+}$ lone pairs lead to strong local polar distortions. The relative orientations of adjacent lone pairs should be opposite to each other and equivalent. The question arises, whether the local polarization is switchable in external electric field (i.e. the system is antiferroelectric) or unswitchable (i.e. just antipolar). Our data support latter possibility, but new experiments with higher electric field are necessary for confirmation of the antipolar order. Note that in both cases the structure remains centrosymmetric at all temperatures.

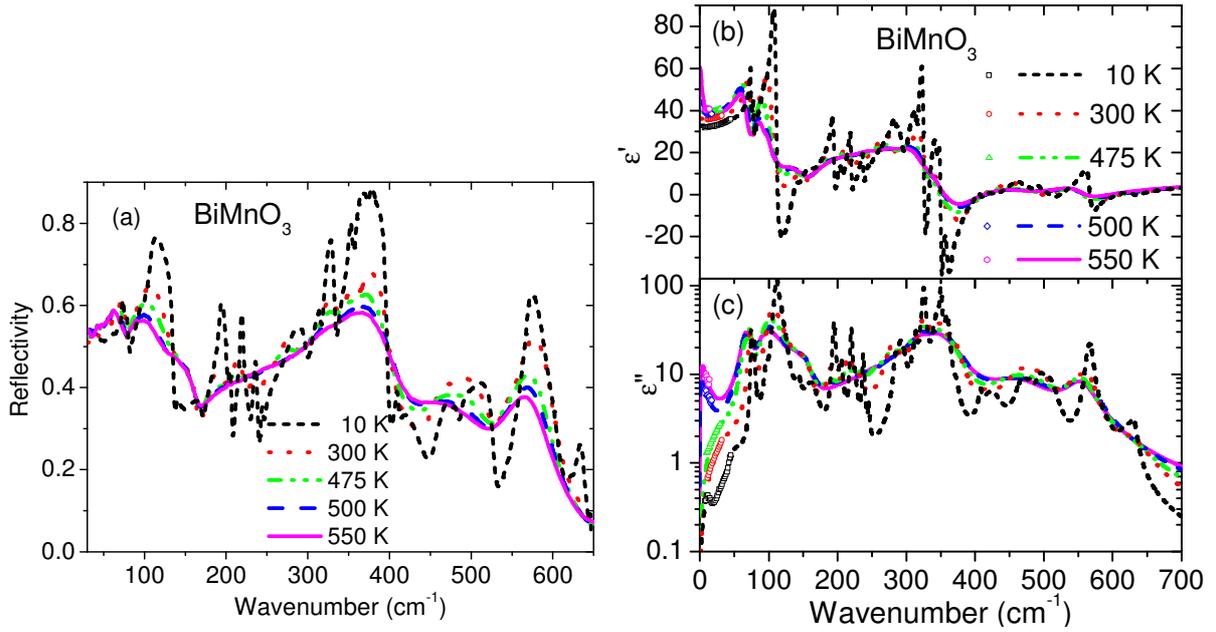

FIG. 2. (Color online) (a) IR reflectivity spectra taken at selected temperatures. (b) Dielectric permittivity $\varepsilon'$ and (c) dielectric loss $\varepsilon''$ spectra obtained from the fits of IR reflectivity and THz spectra. The open symbols seen in (b) and (c) below 50 cm$^{-1}$ are experimental THz dielectric data.

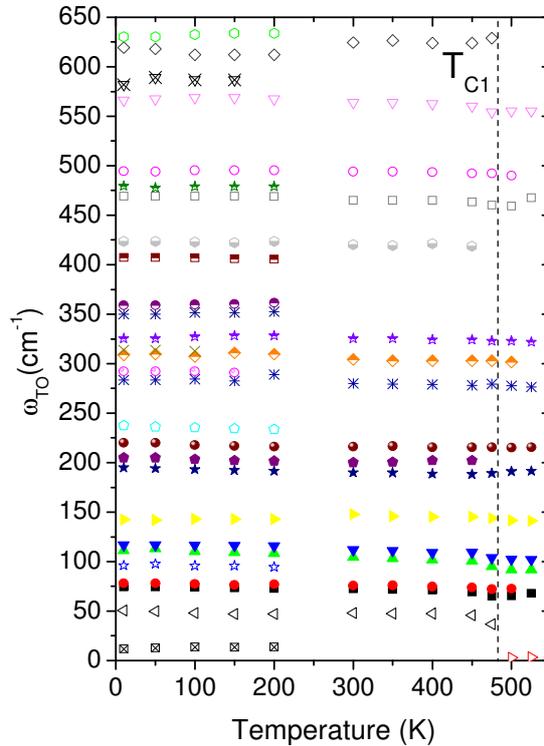

FIG. 3. (Color online) Temperature dependence of selected polar phonon frequencies which exhibit some anomalies near $T_{C1}$ or disappear from the IR spectra on heating. Complete list of phonon frequencies is shown in Table I.

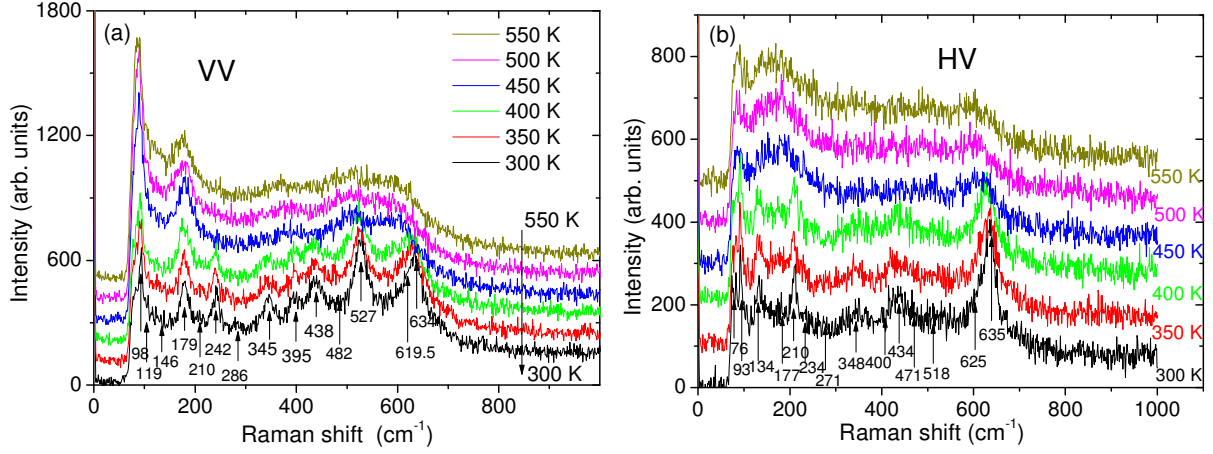

FIG. 4. (Color online) Raman scattering spectra at selected temperatures with (a) parallel (VV) and (b) perpendicular (HV) polarizer and analyzer. Phonon frequencies obtained from the fit of RT Raman spectra are written in the figures.

Raman scattering spectra are shown in Fig. 4. Gradual vanishing of some phonons is seen on heating. This fact will be explained below.

Some polar phonon frequencies exhibit abrupt changes at $T_{C1}$ (Fig. 3). This can be explained by the change of lattice constants observed in XRD.[18] Let us compare the number of observed phonons with the theoretical numbers given by the symmetry of the $BiMnO_3$ system. Early structure refinements gave non-centrosymmetric $C2$-$C_2^3$ space group below $T_{C1}$ with 8 formula units per unit cell.[13,14,15] In such case the factor group analysis of all phonons with wavevector in Brillouin-zone center gives the following:

$$\Gamma_{C2} = 29A(z, x^2, y^2, z^2, xy) + 31B(x, y, yz, xz) \qquad (4)$$

It means that 57 phonons are both IR and Raman active (additional $1A$ and $2B$ modes are acoustic). Recent structural refinements of $BiMnO_3$ incline more toward centrosymmetric $C2/c$ - $C_{2h}^6$ space group.[3,15] In such case we obtain

$$\Gamma_{C2/c} = 14A_g(x^2, y^2, z^2, xy) + 14A_u(z) + 16B_g(xz, yz) + 16B_u(x, y) \ . \qquad (5)$$

It means that 14 $A_g$ and 16 $B_g$ symmetry phonons are only Raman active, while 13 $A_u$ and 14 $B_u$ modes are only IR active (additional 1 $A_u$ and 2 $B_u$ modes are acoustic phonons). We needed 35 oscillators for the fit of 10 K IR reflectivity spectra – see Table 1. It is slightly more than expects $C2/c$ space groups, but simultaneously much less than it is predicted in $C2$ structure. In Raman spectra we observed 15 active modes at room temperature. This is much less than predicted in both above mentioned monoclinic phases. The small number of phonons observed in Raman spectra, is probably caused by high absorption of the sample in visible region (ceramics is black), which markedly reduces Raman scattering volume. Note that the same number of phonons was recently observed in room-temperature Raman spectra of BiMnO$_3$ single crystal.[20]

We believe in *C2/c* structure of BiMnO$_3$, because its factor-group analysis better corresponds to our IR and Raman spectra. The excess of the observed modes in IR spectra can be explained by a) multiphonon absorption (here should be stressed that only summation phonon processes can be seen at low temperatures) or b) by geometrical resonances[26] which can appear in anisotropic media in proximity of two (different symmetry) longitudinal phonon frequencies. The space group as well as number of formula units per unit cell do not change at $T_{C1}$, so no changes of selection rules are expected. Many IR and Raman active phonons gradually disappear on heating, but it is not due to the structural phase transition, but due to gradual increase of the phonon damping and due to overlapping of the phonons. Vanishing of the Raman active modes in Figure 4 at high temperatures can be also caused by increase of conductivity[18] of the sample on heating. In such case, the absorption of the laser beam increases (skin depth drastically decreases) and therefore the Raman scattering volume decreases. Due to this fact the intensity of Raman-active phonons is strongly suppressed at high temperatures and most of the Raman active modes become undetectable.

One can ask what is the origin of the structural phase transition at $T_{C1}$. Based on the analysis of distortions of MnO$_6$ octahedra, it was suggested that the phase transition corresponds to the orbital melting transition.[15] Such transition can cause jump of lattice parameters as well as changes of phonon frequencies observed near $T_{C1}$. Presence of THz dielectric relaxation above $T_{C1}$ gives evidence in dynamical disorder of some atoms. These can be Bi cations, because they are finally responsible for antipolar order below $T_{C1}$. We should admit that there were reported ferroelectric hysteresis loops in thin films,[9,10,11] but applied electric field of 450 kV/cm was much higher than our 20 kV/cm. Such high electric field can significantly influence the crystal structure. One can also speculate about non-stoichiometry of the thin films (Bi is volatile at high temperatures; it is also difficult to control oxygen stoichiometry in the thin films), because the ferroelectric hysteresis loop was observed also in non-stoichiometric bulk samples.[12] Effect of the strain on induction of ferroelectricity in the films can be probably excluded, because the density functional theory calculation found that BiMnO$_3$ stays robustly non-ferroelectric under biaxial strain.[27]

## IV. Conclusion

Our studies did not confirm ferroelectricity in $BiMnO_3$ ceramics. No ferroelectric spontaneous polarization was observed and the number of detected IR and Raman active phonons corresponds better to centrosymmetric *C*2/*c* than to non-controsymmetric *C*2 space group. Small phonon anomalies observed near $T_{C1}$ cannot be explained by earlier reported change of symmetry from *C*2/*c* to *C*2. They are caused by abrupt change of the lattice parameters[18] (without any symmetry change) at $T_{C1}$, which causes the shifts of phonon frequencies. Intrinsic permittivity determined from the THz and IR spectra exhibits abrupt 50% increase at $T_{C1} \approx 475$ K and the saturation at higher temperatures, which reminds antiferroelectric phase transition. Nevertheless, we observed no antiferroelectric hysteresis loop if we applied electric field up to 20 kV/cm. Finally, it can be concluded, that our experimental results support theoretical paper of P. Baettig et al.,[25] who predicted a centrosymmetric crystal structure in $BiMnO_3$ with an antipolar order.

**Acknowledgements**

This work was supported by the Czech Science Foundation (Project No. P204/12/1163) and MŠMT (COST MP0904 project LD12026). A.A.B. acknowledges support from WPI Initiative (MEXT, Japan), JSPS FIRST Program, and JSPS Grant (22246083). In addition, the contribution of F.B. has been supported by the Czech Ministry of Education (Project SVV-2012-265303) and the Czech Science Foundation (Project No. 202/09/H041).

Table 1. Parameters of the polar modes observed at lowest and highest measured temperatures. First mode in the high-temperature phase is a relaxation mode (parameters $\omega_R$ and $\Delta\varepsilon$), the remaining modes are phonons.

| | 10 K | | | 550 K | | |
|---|---|---|---|---|---|---|
| No. | $\omega_{TO}$ (cm$^{-1}$) | $\Delta\varepsilon$ | $\gamma$ (cm$^{-1}$) | $\omega_{TO}$ (cm$^{-1}$) | $\Delta\varepsilon$ | $\gamma$ (cm$^{-1}$) |
| 1 | | | | 3.9 | 24.5 | |
| 2 | 50.5 | 0.2 | 30.4 | 68 | 6.6 | 18.7 |
| 3 | 74.2 | 0.4 | 1.5 | | | |
| 4 | 78 | 1.1 | 3.9 | | | |
| 5 | 95.9 | 0.3 | 4.6 | | | |
| 6 | 111.1 | 7.5 | 8.4 | 90.5 | 1.7 | 15.3 |
| 8 | 116.8 | 1.8 | 9.2 | 102.1 | 3 | 19.7 |
| 9 | 122.8 | 1.4 | 12.5 | 115 | 2.8 | 29.2 |
| 10 | 142.5 | 0.3 | 9.3 | 139.1 | 3.3 | 40.2 |
| 11 | 161.7 | 0.9 | 23.8 | 153.6 | 0.3 | 14 |
| 12 | 194.8 | 0.9 | 4.9 | 200.3 | 0.4 | 44.12 |
| 13 | 204.8 | 0.4 | 6.5 | | | |
| 14 | 219.8 | 0.5 | 3.4 | 222.5 | 0.5 | 41.6 |
| 15 | 225.3 | 0.2 | 7.2 | 249.7 | 0.45 | 36.3 |
| 16 | 237.3 | 0.2 | 3.3 | | | |
| 17 | 245.2 | 0.02 | 2.3 | | | |
| 18 | 283.7 | 0.1 | 5.5 | 280.1 | 1 | 54.3 |
| 19 | 292.2 | 1 | 14.6 | | | |
| 20 | 309 | 0.1 | 5.8 | | | |
| 21 | 313.8 | 0.25 | 4.6 | | | |
| 22 | 325.3 | 1.7 | 6.4 | 320.5 | 3.1 | 55.7 |
| 23 | 344 | 1.2 | 11.4 | 346.4 | 0.8 | 36 |
| 24 | 350 | 0.6 | 4.15 | | | |
| 25 | 359 | 0.8 | 9.3 | | | |
| 26 | 372.8 | 0.04 | 6.5 | 363.2 | 1.6 | 55.83 |
| 27 | 407.2 | 0.04 | 5.8 | | | |
| 28 | 423.5 | 0.4 | 36.5 | | | |
| 29 | 469.2 | 0.35 | 23.6 | 466.2 | 1.7 | 123.03 |
| 30 | 479.5 | 0.02 | 6.8 | | | |
| 31 | 494.5 | 0.2 | 18.1 | | | |
| 32 | 508.4 | 0.3 | 27.2 | | | |
| 33 | 566.2 | 0.4 | 12.6 | 554.2 | 0.5 | 53.6 |
| 34 | 581.8 | 0.1 | 41.6 | | | |
| 35 | 619.2 | 0.05 | 18 | | | |
| 36 | 630 | 0.03 | 10.5 | | | |